%% file: main.tex
\documentclass{article}

\usepackage[final]{neurips_2024_ml4ps}

\usepackage[utf8]{inputenc} 
\usepackage[T1]{fontenc}    
\usepackage{hyperref}       
\usepackage{url}            
\usepackage{booktabs}       
\usepackage{amsfonts}       
\usepackage{nicefrac}       
\usepackage{microtype}     
\usepackage{xcolor}         
\usepackage{import}
\usepackage{amsmath}
\usepackage{amsthm}
\usepackage{graphicx}
\usepackage{cancel}
\usepackage{subcaption}
\usepackage{float}
\usepackage{wrapfig}
\usepackage{verbatim}
\usepackage{hyperref}
\usepackage{mwe}
\usepackage{wrapfig}
\usepackage{algorithm}
\usepackage{algpseudocode}
\usepackage{multicol}
\usepackage{quiver}
\usepackage{adjustbox}

\definecolor{alessiogreen}{RGB}{0, 192, 0}

\newcommand{\mycomment}[1]{}

\title{Bayesian Deconvolution of Astronomical Images with Diffusion Models: Quantifying Prior-Driven Features in Reconstructions
}

\author{%
Alessio Spagnoletti$^{1,2}$ \quad Alexandre Boucaud$^{1}$ \quad Marc Huertas-Company$^{3}$ \\
\textbf{Wassim Kabalan}$^{1}$ \quad \textbf{Biswajit Biswas}$^{1}$ \\
$^1$Université Paris Cité, CNRS, Astroparticule et Cosmologie, F-75013 Paris, France \\
$^2$Université Paris Cité, CNRS, MAP5, F-75006 Paris, France \\
$^3$Instituto de Astrofísica de Canarias (IAC), Canarias, Spain \\
\texttt{\{alessio.spagnoletti, alexandre.boucaud, wassim.kabalan,}\\\texttt{ biswajit.biswas\}@apc.in2p3.fr}\\
\texttt{mhuertas@iac.es}
}

\begin{document}
\maketitle

\begin{abstract}
  Deconvolution of astronomical images is a key aspect of recovering the intrinsic properties of celestial objects, especially when considering ground-based observations. This paper explores the use of diffusion models (DMs) and the Diffusion Posterior Sampling (DPS) algorithm to solve this inverse problem task. We apply score-based DMs trained on high-resolution cosmological simulations, through a Bayesian setting to compute a posterior distribution given the observations available. By considering the redshift and the pixel scale as parameters of our inverse problem, the tool can be easily adapted to any dataset. We test our model on Hyper Supreme Camera (HSC) data and show that we reach resolutions comparable to those obtained by Hubble Space Telescope (HST) images. Most importantly, we quantify the uncertainty of reconstructions and propose a metric to identify prior-driven features in the reconstructed images, which is key in view of applying these methods for scientific purposes.
\end{abstract}

\import{./}{Introduction}

\import{./}{Diffusion_Models}

\import{./}{Experiments}

\import{./}{Conclusion}

\bibliographystyle{abbrv}
\bibliography{biblio}

\appendix

\import{./}{Hyperparameters}

\import{./}{UMAP}

\import{./}{Images}

\end{document}

%% file: Introduction.tex
\section{Introduction}

The analysis of astronomical observations is affected by multiple sources of degradation and noise, such as the point spread function (PSF) due to atmospheric interference and optical effects \cite{Starck2002DeconvolutionIA}. Although widely used, traditional deconvolution methods often encounter significant limitations, including noise amplification, loss of resolution, and sensitivity to model assumptions \cite{Lucy1994PSF2,Bertero1998PSF3,Starck2002DeconvolutionIA,Michalewicz2023PSF1}.

In this paper, we explore and evaluate the application of diffusion models (DMs) \cite{Song2019, Song2020DenoisingDI, Song2020improved, Song2021SDE, Ho2020Denoising, Nichol2021ImprovedDDPM}, for astronomical image deconvolution, through the diffusion posterior sampling (DPS) algorithm \cite{Chung2022}. DMs have gained attention in recent years for their ability to model complex data distributions through iterative denoising processes, often outperforming generative adversarial networks in image generation tasks \cite{Dhariwal2021}. They provide a flexible and powerful approach to image restoration, either through the use of plug-and-play algorithms \cite{Venkatakrishnan2013PlugP, Graikos2022DiffusionPlugP} or by directly conditioning the diffusion process using a log-likelihood guidance term, $\nabla_{\mathbf{x}} \log p(\mathbf{y}|\mathbf{x}_t)$ \cite{Song2021SDE}. DPS follows this latter approach.

The objective is to leverage realistic simulations to train a score-based diffusion model that can be applied for deconvolution and noise removal on real images. We choose TNG100 simulations from the TNG Illustris project \cite{Nelson2018TNG, Bottrell2023IllustrisTNGHSC} to train our DM and HSC-PDR3 Survey \cite{Aihara2021HSC} data to test the DPS.

We highlight the pros and cons of our approach, examining factors such as image quality, computational efficiency, robustness to noise, and tendency to hallucinate \cite{Liu2007FaceHall, Wang2013FaceHall2, Sampson2023SpottingHI}. While our preliminary results suggest that our framework holds promise for improving deconvolution outcomes, we also identify several areas where the model struggles, particularly in keeping physical consistency and achieving real-time processing speeds.

%% file: Diffusion_Models.tex
\section{Model and Datasets}\label{sec:Diffusion}

\paragraph{Score-based diffusion model}

We follow the variance-preserving stochastic differential equation (VP-SDE) approach from \cite{Song2021SDE}, focusing on the Denoising Diffusion Probabilistic Model (DDPM) framework \cite{Song2020DenoisingDI}. The general idea is to define the forward process $\mathcal{F}$ that perturbs the original data $\mathbf{x}_0$ into a noisy version of itself $\mathbf{x}_t$ at time $t$, and a backward process $\mathcal{B}$ that solves a reverse-time stochastic process \cite{Anderson1982Reversetime}, written as:
\footnotesize
\begin{equation*}
    \mathcal{F}: \ d\mathbf{x} = - \frac{\beta(t)}{2} \mathbf{x} \, dt + \sqrt{\beta(t)} \, d\mathbf{w}, \quad \quad
    \mathcal{B}: \ d\mathbf{x} = \left( - \frac{\beta(t)}{2} \mathbf{x} - \beta(t) \nabla_{\mathbf{x}_t} \log p_t(\mathbf{x}_t) \right) dt + \sqrt{\beta(t)} \, d\bar{\mathbf{w}}
\end{equation*}
\normalsize
where $\beta(t)$ controls how much variance is added at each step, $\mathbf{w}$ is standard Wiener process and $\bar{\mathbf{w}}$ is a backward Wiener process defined in \cite{Anderson1982Reversetime}.

When conditioning on an observation $\mathbf{y}$, the posterior term $p( \mathbf{x}_t | \mathbf{y})$ is introduced to guide the reverse process \cite{Song2021SDE}.  Applying Bayes' rule yields
\footnotesize
\begin{equation}\label{eq:Bayes}
    d\mathbf{x} = \left( - \frac{\beta(t)}{2} \mathbf{x} - \beta(t) \left[ \nabla_{\mathbf{x}_t} \log p_t(\mathbf{x}_t) + \nabla_{\mathbf{x}_t} \log p_t(\mathbf{y} | \mathbf{x}_t) \right] \right) dt + \sqrt{\beta(t)} \, d\bar{\mathbf{w}}\,.
\end{equation}
\normalsize
The DPS algorithm approximates the likelihood $p(\mathbf{y} | \mathbf{x}_t)$
(intractable analytically \cite{Song2021SDE, Chung2022}) with $p(\mathbf{y} | \hat{\mathbf{x}}_0)$, where $\hat{\mathbf{x}}_0$ is an estimate of the recovered data. Thanks to Tweedie’s formula \cite{Robbins1956AnEB, Stein1981EstimationOT, Efron2011TweediesFA, Kim2021Noise2ScoreTA}, using the score of the prior learned during the training process $s_\theta^* \simeq \nabla_{\mathbf{x}_t} \log p_t(\mathbf{x}_t)$ and a cumulative noise schedule $\bar{\alpha}(t)$ that depends on $\beta(t)$ \cite{Song2021SDE}, its expression is
\footnotesize
\begin{equation*}
\hat{\mathbf{x}}_0 \simeq \frac{1}{\sqrt{\bar{\alpha}(t)}} \left[ \mathbf{x}_t + \left( 1 - \bar{\alpha}(t) \right) s_\theta^* (\mathbf{x}_t, t) \right]\,.
\end{equation*}
\normalsize

Our publicly available DDPM and DPS implementation\footnote{\url{https://github.com/astrodeepnet/diffusion4astro.git}} is based on \cite{Nichol2021ImprovedDDPM} and \cite{Chung2022}, respectively, for the training phase and deployment. Implementation details can be found in Appendix \ref{sec:hyperpar}.

\paragraph{TNG Illustris simulations}
The selected TNG100 simulated images have been made from snapshots of the evolution of the simulated universe at specific redshifts $z = 0.1527,\,0.1693$ and $0.1804$ with camera field-of-view face-on (also referred to as \texttt{v0} in fits files). 
These snapshots have a fixed physical resolution of $0.4$\,kpc/pixel. The images are cropped at center to a size of 256 pixels, and we consider only the {\em g,r} and {\em i} bands. We refer to these images as the {\em idealized} TNG. This dataset is composed of $17852$ images, that is split between a training and test with a $0.9/0.1$ ratio. 
To increase the dynamic range, we normalize the images and use an \texttt{asinh} stretch with a scale of $0.01$.

To deconvolve real images $\mathbf{y}$ from the HSC survey, we define the operator $\mathcal{A}$ for which we assume that the observation is defined as $\mathbf{y} = \mathcal{A}(\mathbf{x}) + \sigma_{\mathbf{y}}\mathbf{n}$ assuming $\mathbf{n}$ a standard Gaussian noise, and $\sigma_{\mathbf{y}}$ is the noise variance. Our goal is thus to recover $\mathbf{x}$.

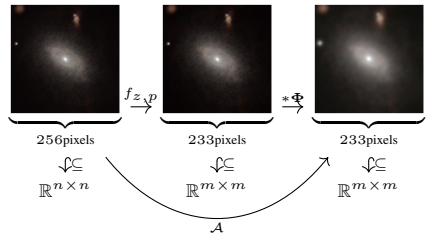
\begin{wrapfigure}{r}{.41\textwidth}
    \vspace{-.9cm}
    \begin{equation*}
        \scriptsize
        \begin{tikzcd}[column sep=small, row sep=small]
            {\underbrace{\includegraphics[width=1.5cm]{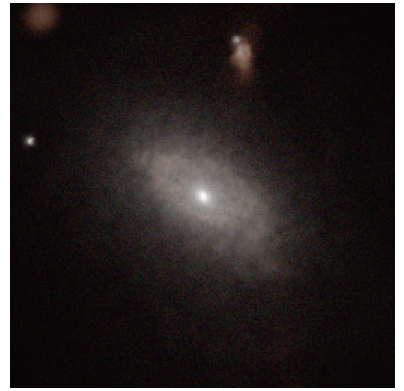}}_{256 \text{pixels}}} 
            & {\underbrace{\includegraphics[width=1.5cm]{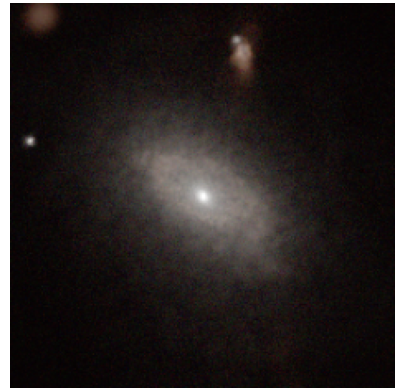}}_{233 \text{pixels}}} 
            & {\underbrace{\includegraphics[width=1.5cm]{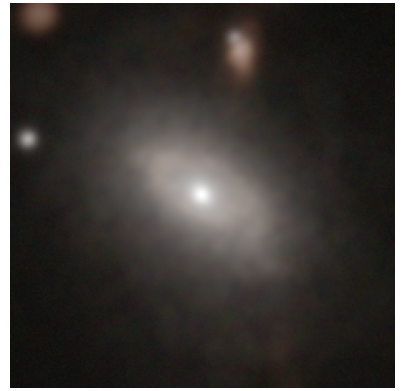}}_{233 \text{pixels}}} \\
            {\mathbb{R}^{n \times n}} & {\mathbb{R}^{m \times m}} & {\mathbb{R}^{m \times m}}
            \arrow["{f_{z,p}}", from=1-1, to=1-2]
            \arrow["{\mathcal{A}}"', curve={height=50pt}, from=1-1, to=1-3]
            \arrow[hook, "\subseteq", from=1-1, to=2-1]
            \arrow["{*\mathbf{\Phi}}", from=1-2, to=1-3]
            \arrow[hook, "\subseteq", from=1-2, to=2-2]
            \arrow[hook, "\subseteq", from=1-3, to=2-3]
        \end{tikzcd}
        \vspace{-.3cm}
    \end{equation*}
    \scriptsize
    \caption{Schematic of the process to transform an idealized TNG simulated image into a HSC-like observation through the operator $\mathcal{A}$.}
    \normalsize
    \vspace{-.7cm}
    \label{fig:pipeline}
\end{wrapfigure}

Due to projection effects, the apparent size of galaxies (i.e. the size to which they appear to the camera) at various redshifts is different; 
Since the model is trained with a pixel scale of fixed physical units per pixel (the idealized TNG data), one must then incorporate in the operator $\mathcal{A}$ a redshift-dependent scaling 
to match their apparent size. Furthermore, the matching of HSC pixel scale $p=0.17$ arcsec/pixel must be done before the images are convolved with the HSC PSF $\mathbf{\Phi}$ to simulate an observation. In Figure \ref{fig:pipeline}, we produce a schematic of that process, where the matching operations are grouped under the $f_{z,p}$ function, and $m := m(z,p)$ is the size of the observed image.

\paragraph{HSC - PDR3 Survey data}
The Hyper Suprime-Cam Subaru Strategic Program (HSC-SSP)\footnote{Official website: \url{https://hsc.mtk.nao.ac.jp/ssp/}} is an astronomical survey conducted using the Hyper Suprime-Cam (HSC) on the Subaru Telescope, a powerful 8.2-meter optical telescope located in Hawaii. The data from this survey, particularly from its third public data release (PDR3) \cite{Aihara2021HSC}, has been widely used in various research fields. We fetch HSC images along with their approximated PSF\footnote{PSF Picker tool: \url{https://hsc-release.mtk.nao.ac.jp/psf/pdr3/}} and computed photometric redshift \cite{Tanaka2017Photometric1, Nishizawa2020Photometric2, Nishizawa2022Photometric3}.

%% file: Experiments.tex
\section{Experiments}

The training process for the model required a total of \textbf{133 hours} on \textbf{32 V100 GPUs}. Following $300k$ training steps, an analysis revealed that the model exhibited a tendency toward mode collapse \cite{Goodfellow2016}. This was identified by projecting the training dataset and the model predictions onto a lower-dimensional manifold using UMAP \cite{McInnes2018UMAP} (refer to Appendix \ref{sec:UMAP} for corresponding visualizations). To mitigate this issue, the training was extended by an additional $225k$ steps, accompanied by an increase in the batch size from $64$ to $128$ and the introduction of a random shift of up to 10 pixels in any direction for the training images. This randomness enabled the model to generate galaxies that were not perfectly centered, aligning more closely with real-world observations where object centers are often uncertain.

To qualify the performance of our model, we computed the Fréchet Inception Distance (FID) \cite{Heusel2017FID}, a metric that indicates how well the generated samples match the features of the training images. The values computed on $1024$ images sampled from our models are shown on Table~\ref{tab:FID}. The slight increase between $450k$ and $525k$ may be explained by the random shift, which was not present in the training set, for which all galaxies were perfectly centered. Since our work is the first to use this specific part of the TNG Illustris dataset, we cannot compare these numbers with others, but they may be helpful to set a benchmark future works.
\begin{table}[h!]
    \centering
    \begin{tabular}{l|c|c|c|c}
    
     \textbf{Model training steps} & 300k & 375k & 450k & 525k \\ \hline
    \textbf{FID value} $\downarrow$   & 317.3 & 235.3 & \textbf{119.4} & 132.6  \\ 
    
    \end{tabular}
    \vspace{0.5em}
    \caption{Performance of our model as a function of training steps (lower is better).}
    \label{tab:FID}
    \vspace{-0.65cm}
\end{table}
\vspace{-0.1cm}
\subsection{Results on simulated observations}

Since our model has only been trained on simulated images, a first measure of the performance of our algorithm is to deconvolve a simulated TNG image that has previously been convolved with a realistic PSF to simulate an observation. This is possible due to the publicly available TNG dataset \cite{Bottrell2023IllustrisTNGHSC} which not only provides the simulations we used to train our model but also contains HSC-like images. These are indeed simulated TNG images that have been convolved with HSC PSF, set to the HSC pixel scale, and put at a specific location in the HSC-SSP survey, providing a real background to the simulated objects.

\begin{figure}[H]
    \centering
    \begin{minipage}{0.73\textwidth}
        \centering
        \includegraphics[width=\linewidth]{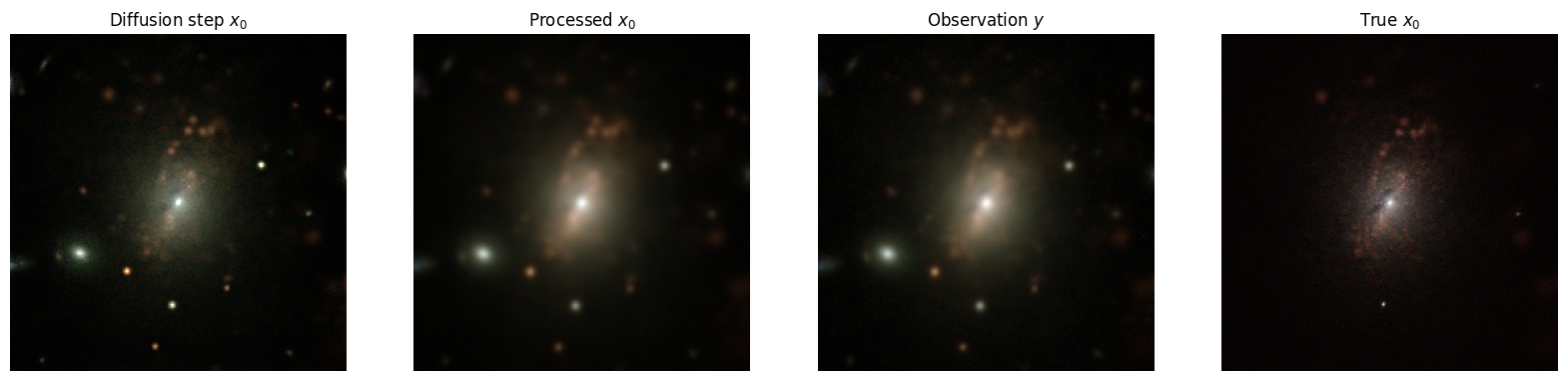}
        \caption{Comparison of DDPM TNG-HSC deconvolved image (leftmost) with \textit{Idealized} TNG counterpart (rightmost). $z=0.1768$}
        \label{fig:ExpTNG_res1}
    \end{minipage}
    \hfill
    \begin{minipage}{0.2\textwidth}
        \centering
        \vspace{-0.32cm}
        \includegraphics[width=0.8\linewidth]{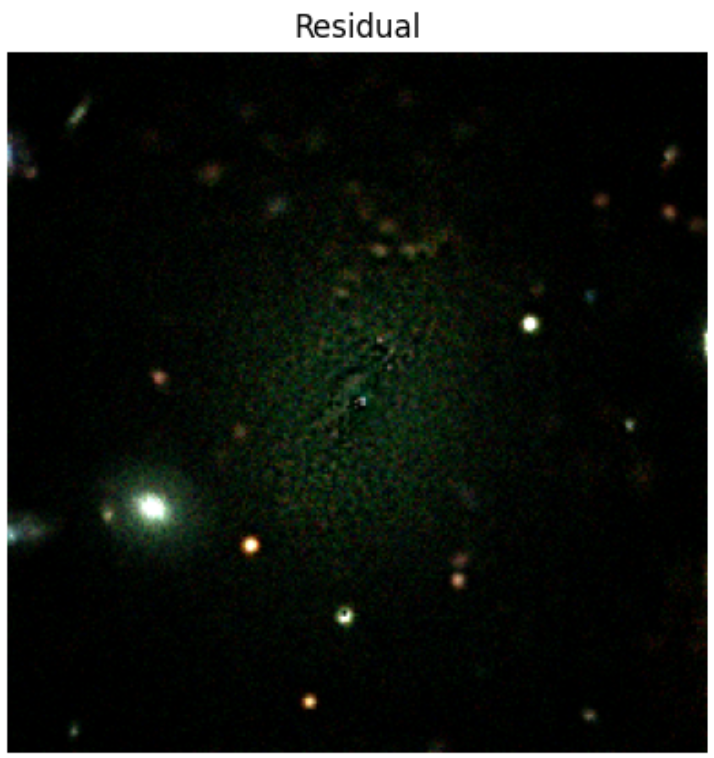}
        \vspace{0.04cm}
        \caption{Residual}
        \label{fig:residuals}
    \end{minipage}
\end{figure}

In Figure \ref{fig:ExpTNG_res1} we show respectively: the final step of the diffusion process $\mathbf{x}_0^{\text{Diff}}$, the re-convolved final step that can be compared to the TNG-HSC observation $\mathbf{y}$, and the true TNG $\mathbf{x}_0^{\text{Ideal}}$. Both the TNG-HSC and TNG images are stretched with \texttt{asinh} and scale 0.01. We notice how the re-convolved final step image, i.e. $\mathcal{A}(\mathbf{x}_0^{\text{Diff}})$, is close to the observation $\mathbf{y}$, which is an empirical proof that the reconstruction process did not hallucinate by creating new information. In Figure \ref{fig:residuals} we show the residual between the reconstructed $\mathbf{x}_0^{\text{Diff}}$ and the $\mathbf{x}_0^{\text{Ideal}}$, from which we can see that the galaxy is well reconstructed since most residuals are due to the artificially added HSC background.

\subsection{Comparing with space-based imaging}

Thanks to the data publicly available through the PDR3 Survey UltraDeep layer \cite{Aihara2021HSC}, we can crop areas\footnote{Visual tool: \url{https://hsc-release.mtk.nao.ac.jp/hscMap-pdr3/app}} at known astronomical coordinates to retrieve the \textit{gri} bands, the associated PSFs and photometric redshifts. We chose an object in a location in the sky where we both have images from the COSMOS project \cite{Scoville2006COSMOS} and PDR3. We then apply the DPS algorithm with our DDPM model trained on \textit{idealized} TNG images and compare the result with the equivalent Hubble Space Telescope (HST - ACS) images, downsampled at the HSC pixel-scale. The HST is only taken at channel \textit{i}, i.e., the F814W band, to provide a visual comparison.

\begin{figure}[htbp]
    \centering
    \begin{minipage}{0.55\textwidth}
        \centering
        \includegraphics[width=\linewidth]{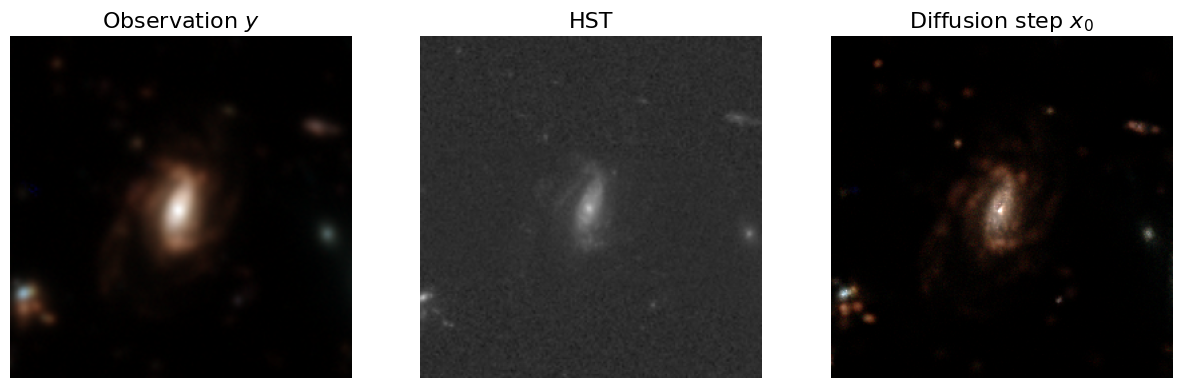}
        \caption{Comparison of DDPM HSC deconvolved image (rightmost) with HST counterpart (middle).\\Ra-dec coords (10h1m49.30s, 2°23'18.60") $z=0.2168$}
        \label{fig:Exp2_res1}
    \end{minipage}
    \hfill
    \begin{minipage}{0.4\textwidth}
        \centering
        \begin{subfigure}[b]{0.45\textwidth}
            \centering
            \includegraphics[width=0.925\linewidth]{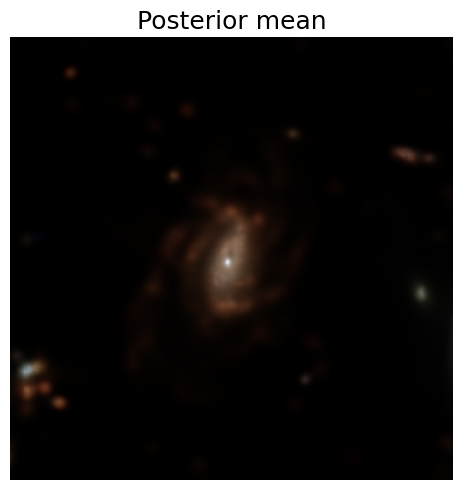}
            \vspace{0.01cm}
        \end{subfigure}
        \hfill
        \begin{subfigure}[b]{0.5\textwidth}
            \centering
            \includegraphics[width=\linewidth]{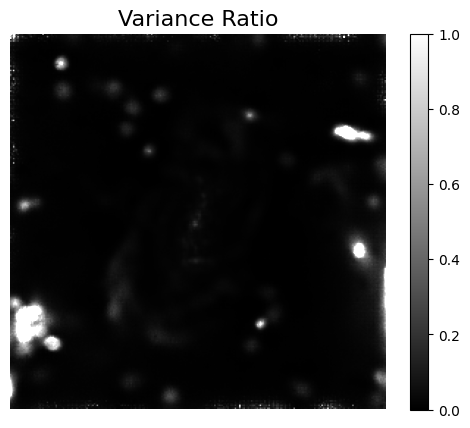}
            \vspace{-0.43cm}
        \end{subfigure}
        \caption{Posterior mean (left) and variance-ratio (right) of pixel values over 256 samples from Figure~\ref{fig:Exp2_res1}.}
        \label{fig:test_avg_and_var}
    \end{minipage}
\end{figure}

\begin{figure}[htbp]
    \centering
    \begin{minipage}{0.55\textwidth}
        \centering
        \includegraphics[width=\linewidth]{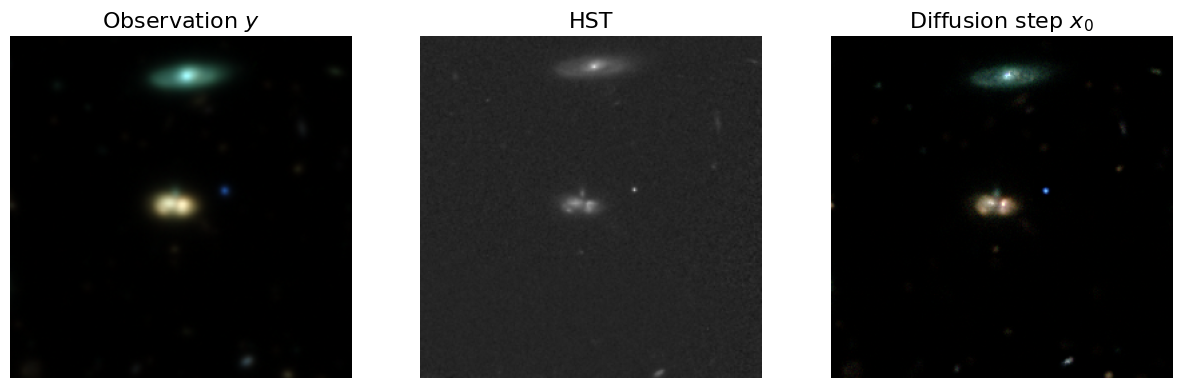}
        \caption{Comparison of DDPM HSC deconvolved image (rightmost) with HST counterpart (middle).\\Ra-dec coords (10h 1m 34.77s, 2°4'19.02") $z=0.1665$}
        \label{fig:Exp2_res3}
    \end{minipage}
    \hfill
    \begin{minipage}{0.4\textwidth}
        \centering
        \begin{subfigure}[b]{0.45\textwidth}
            \centering
            \includegraphics[width=0.925\linewidth]{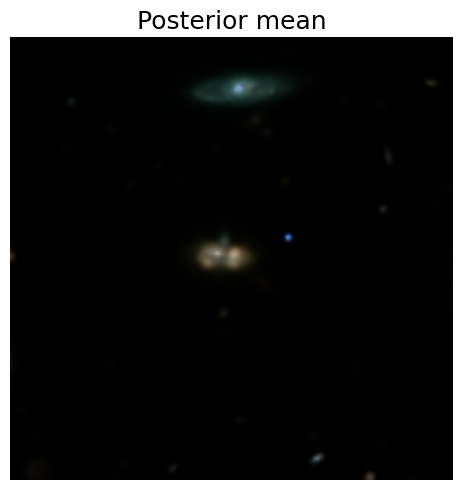}
            \vspace{0.01cm}
        \end{subfigure}
        \hfill
        \begin{subfigure}[b]{0.5\textwidth}
            \centering
            \includegraphics[width=\linewidth]{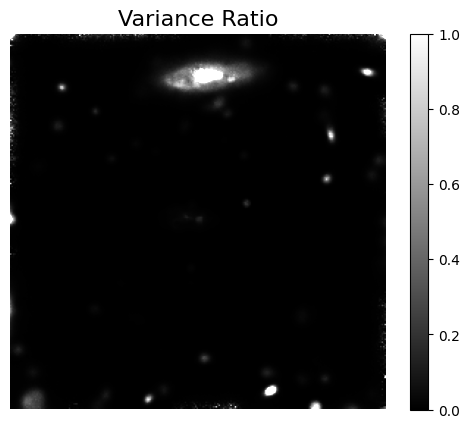}
            \vspace{-0.43cm}
        \end{subfigure}
        \caption{Posterior mean (left) and variance-ratio (right) of pixel values over 256 samples from Figure~\ref{fig:Exp2_res3}.}
        \label{fig:test_avg_and_var_3}
    \end{minipage}
\end{figure}

In Figure \ref{fig:Exp2_res1} and  \ref{fig:Exp2_res3}, we show the ground-based observation (HSC) $\mathbf{y}$, the space-based reference (HST), and the final step of the diffusion process $\mathbf{x}_0$. Both the HSC and HST images are stretched with \texttt{asinh} and scale 0.01. Each evaluation of the DPS takes $\mathbf{\sim100}$ \textbf{seconds} on \textbf{a single A100 GPU}. Indeed, the time consumption is a negative aspect of the DPS, which doubles the DDPM sampling time due to the extra computation of the log-likelihood gradient term in Equation~\eqref{eq:Bayes}.
\vspace{-0.1cm}
\vspace{-0.7cm}

\vspace{-0.1cm}

To provide a more insightful analysis, we sample $256$ images from the posterior distribution and compute the pixel-wise mean and variance (ratio). Figures~\ref{fig:test_avg_and_var}, \ref{fig:test_avg_and_var_3} show that the posterior mean represents a reliable output for the inverse problem since it averages out the stochastic fluctuations. On the right, the variance ratio image is produced by dividing the posterior variance by the prior variance, giving an idea of the areas in which the uncertainty on the generated image is higher, i.e., the prior-dominated areas.
The fact that the central part of the variance ratio image is close to 0, means that the model is not hallucinating when reconstructing the galaxy. However the reconstructed outskirts of the galaxy seem to be much more prior-dominated, probably due to the nature of the TNG training dataset.

\subsection{Limitations}

\begin{figure}[htbp]
    \centering
    \begin{minipage}{0.55\textwidth}
        \centering
        \includegraphics[width=\linewidth]{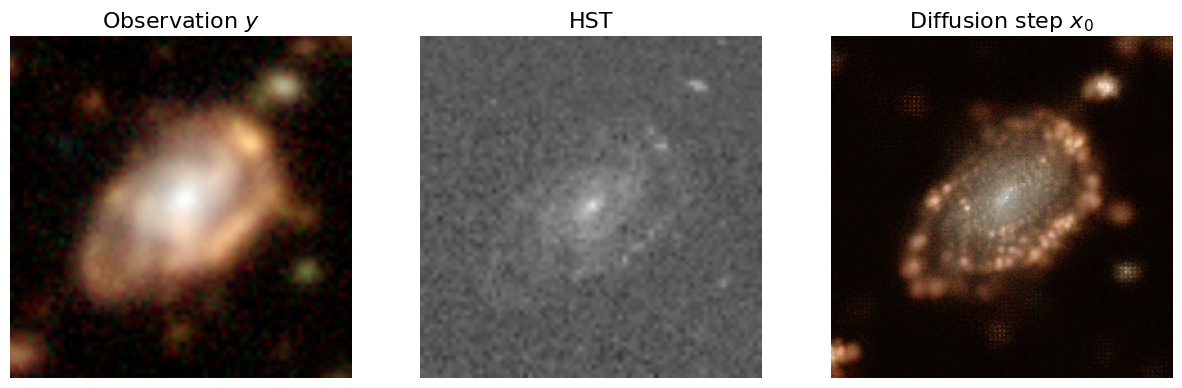}
        \caption{Comparison of DDPM HSC deconvolved image (rightmost) with HST counterpart (middle).\\Ra-dec coords (10h0m2.86s, 2°2'13.92") $z= 0.8463$.}
        \label{fig:exp2_fail}
    \end{minipage}
    \hfill
    \begin{minipage}{0.4\textwidth}
        \centering
        \begin{subfigure}[b]{0.45\textwidth}
            \centering
            \includegraphics[width=0.925\linewidth]{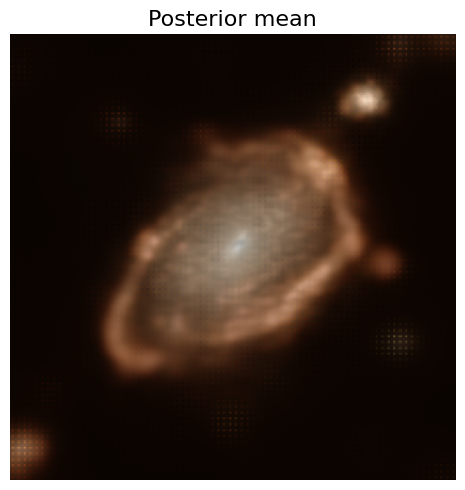}
            \vspace{0.01cm}
        \end{subfigure}
        \hfill
        \begin{subfigure}[b]{0.5\textwidth}
            \centering
            \includegraphics[width=\linewidth]{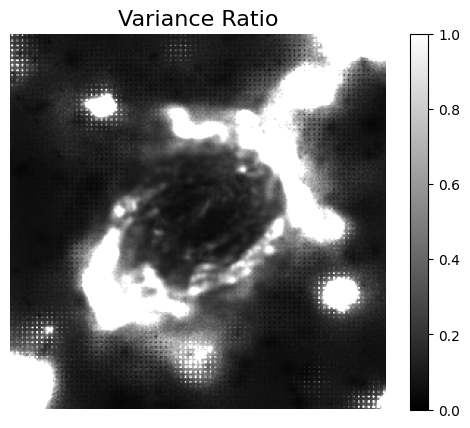}
            \vspace{-0.43cm}
        \end{subfigure}
        \caption{Posterior mean (left) and variance-ratio (right) of pixel values over 256 samples from Figure~\ref{fig:exp2_fail}.}
        \label{fig:fail_avg_and_var}
    \end{minipage}
    \vspace{-0.3cm}
\end{figure}

Despite the good results, the model has limitations when the redshift $z$ gets too high. Indeed, at $z>0.7$, we get images of size $m(z,p)<80$. This means that when we try to solve the inverse problem, we are both solving the deconvolution and the upsampling tasks, and an upscaling factor higher than a factor $2$ is problematic. We notice on Figure \ref{fig:exp2_fail} how a single sampled galaxy at $z>0.8$ is highly prior-dominated. In particular, a high number of features are created (hallucinated) all around the galaxy, and are invisible on the HST counterpart. This is common when the solution to the inverse problem requires injecting too much information from the prior (TNG idealized images).

Considering again $256$ posterior samples for this galaxy and computing the mean and variance ratio, demonstrate this prior domination on a large part of the image on Figure \ref{fig:fail_avg_and_var}. The posterior mean is still quite robust with fewer small TNG features than a single sample.

\subsection{Quantification of hallucinations}

To better qualify the impact of the prior on the deconvolution, we define a metric based on the posterior variance. Building on the results we explored on Figures~\ref{fig:test_avg_and_var}, \ref{fig:test_avg_and_var_3} and \ref{fig:fail_avg_and_var}, we take the pixel-wise posterior variance and prior variance over 256 samples, take their ratio and compute the average value on a centered aperture or radius $r$ (here we chose $r=30$ pixels) that contains the central part of the galaxy. 
To verify such metric is positively correlated with a prior-dominated reconstruction, we create a test using simulated TNG images, which we artificially convolve with a Gaussian PSF and add a fixed noise level $\sigma_{\mathbf{y}}$ measured on HSC images. The main point is to vary the signal-to-noise ratio (SNR) by setting specific magnitudes to the galaxy (it will be lower when the magnitude is higher, i.e. the galaxy fainter). We vary the magnitude from $18$ to $23$, and report the metric results in Table~\ref{tab:hall}, where we confirm such positive correlation when the noise starts dominating the galaxy signal. 

We refer to Appendix \ref{sec:imag} for the visual results at all magnitudes.

\begin{table}[h!]
    \centering
    \begin{tabular}{c|c|c|c|c|c|c}
    
    \textbf{Simulated magnitude} & 18 & 19 & 20 & 21 & 22& 23\\ \hline
    \textbf{SNR} & 699.3 & 278.4 & 110.8 & 44.1 & 17.6 & 7.0\\
    \hline
    \textbf{Variance Ratio} ($r=30$)  & $0.0114$ & $0.0116$ & $0.0168$ & $0.0346$ & $0.0560$ & $0.0685$ \\ 
    
    \end{tabular}
    \vspace{0.5em}
    \caption{Signal-to-noise and variance ratio as a function of the simulated TNG galaxy magnitude.}
    \label{tab:hall}
\end{table}
\vspace{-.5cm}

%% file: Conclusion.tex
\section{Conclusion}

By exploring diffusion-based methods in the context of astronomical image deconvolution, we aim to provide a foundation for future research and development of signal processing in astronomy. Our findings contribute to the growing body of works on applying advanced machine learning techniques to astrophysical data analysis \cite{Lanusse2020DeepGM, Ntampaka2019TheRoleof, Adam2022GalaxyPosterior, SmithDDPMgal}. In particular, one of our inspirations comes from the work of \cite{adam2023echoesnoiseposteriorsamples}, who first implemented DMs to tackle the astronomical deconvolution problem. Our work builds on it by including the redshift and pixel scale as parameters of the inverse problem to adapt the model to any dataset. We also propose an implementation of a general-purpose algorithm, the DPS, to solve the deconvolution of astronomical observations and make it publicly available, allowing the reproducibility of the results. Finally, we provide a metric to identify prior-driven features in the generated solutions. We hope to take a step toward providing confidence to the scientific community for applying these models to real-case scenarios.

\section*{Broader impact}

Generative modelling for improving data quality in astrophysics has been investigated quite extensively over the past decade \cite{Schawinski_2017, Vojtekova2020, Lauritsen2021, Gan_2021}. However, despite promising results, very few works use these techniques for scientific applications because it is difficult to identify whether the model outputs can be trusted. In this work, we try to go a step forward towards deployment by investigating ways to track model hallucinations and prior-driven features. This opens the door to a systematic application of these deconvolution methods on large-scale statistical samples of celestial objects to infer physical properties. We plan to investigate the use of score-based deconvolution techniques to resolve the internal structure of galaxies. 

\section*{Acknowledgments and Disclosure of Funding}

This work was granted access to the HPC resources of IDRIS under the allocation 2024-AD011014557 made by GENCI. This research was supported by the Data Intelligence Institute of Paris (diiP), and IdEx Université de Paris (ANR-18-IDEX-0001). AS acknowledges support from the France 2030 research programme on artificial intelligence, via PEPR PDE-AI grant (ANR-23-PEIA-0004), and from the French Research Agency through the PostProdLEAP project (ANR-19- CE23- 0027-01). BB acknowledges funding from the European Union’s Horizon 2020 research and innovation program under the Marie Skłodowska-Curie grant agreement No 945304 – COFUND AI4theSciences hosted by PSL University, and Agence Nationale de la Recherche grant ANR-19-CE23-0024 — AstroDeep.

%% file: Hyperparameters.tex
\section{Training details}\label{sec:hyperpar}

The model used to approximate the score function introduced in Section~\ref{sec:Diffusion} is the UNet \cite{Ronneberger2015UNet} model defined in \cite{Nichol2021ImprovedDDPM}. To train the network we used the following hyperparameters:

\begin{itemize}
    \item \texttt{--image\_size 256}
    \item \texttt{--num\_channels 192}
    This flag indicates the base number of channels used in the Unet's convolutional layers.

    \item \texttt{--num\_res\_blocks 3}
    This specifies the number of \textbf{residual blocks} to use at each level in the UNet architecture.

    \item \texttt{--diffusion\_steps 1000}

    \item \texttt{--noise\_schedule cosine}
    The model uses a \textbf{cosine noise schedule}. The cosine schedule has been shown \cite{Nichol2021ImprovedDDPM} to improve model performance by making the noise added to the images more evenly distributed across the steps (as opposed to the traditional linear schedule).

    \item \texttt{--lr 1e-4}
    This flag specifies the learning rate.

    \item \texttt{--batch\_size 64 or 128}
    This flag specifies the batch size used during training. We train our model for $300k$ steps using a batch size of $64$, and then we improve the results by extending the training to $525k$ steps using a batch size of $128$.
\end{itemize}

%% file: UMAP.tex
\section{UMAP representation of the data}\label{sec:UMAP} 

We use the low-dimensional manifold projection called UMAP~\cite{McInnes2018UMAP} to visualize on the same manifold the data distribution from the training set (blue dots) and the one sampled from the generative model (red dots). We use the fact that the distribution of the sampled data is not as homogeneous as the one from the training as a sign that the training is incomplete or collapses into specific modes as shown in Figure~(a). After additional training, the behavior improves as seen on Figure~(b).
\vspace{1cm}
\begin{figure}[htbp]
    \centering
    \begin{minipage}{\textwidth}
        \centering
        \begin{subfigure}[b]{0.45\textwidth}
            \centering
            \includegraphics[width=0.9\linewidth]{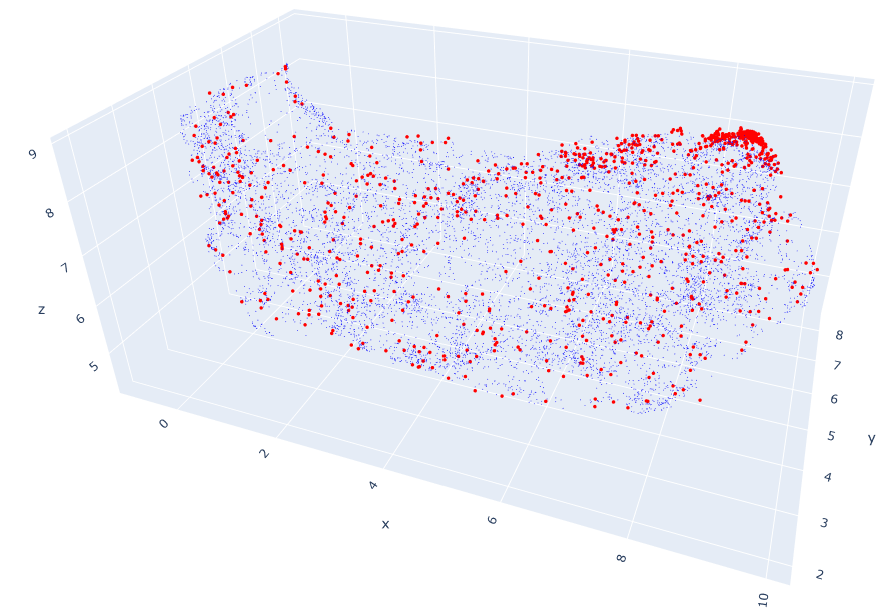}
            \caption{UMAP 3D representation of the training dataset (\color{blue}blue dots\color{black}), as well as 1000 samples (\color{red}red dots\color{black}) generated from the model after $300k$ steps. We can spot a mode collapse at the top right.}
            \label{fig:collapse}
        \end{subfigure}
        \hfill
        \begin{subfigure}[b]{0.45\textwidth}
            \centering
            \includegraphics[width=0.9\linewidth]{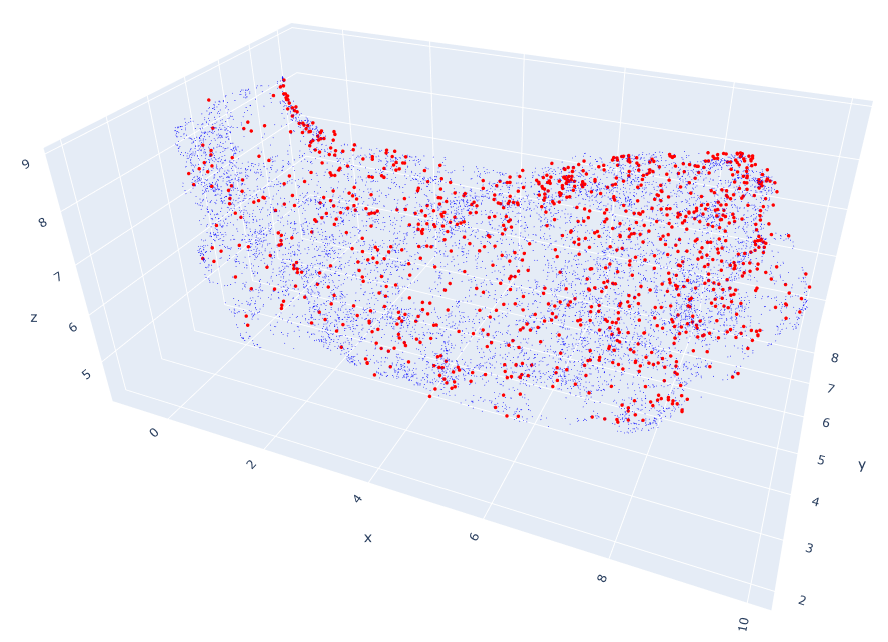}
            \caption{UMAP 3D representation of the training dataset (\color{blue}blue dots\color{black}), as well as 1000 samples (\color{red}red dots\color{black}) generated from the model after $525k$ steps. Much more homogeneous.}
        \end{subfigure}
    \end{minipage}
    \label{fig:UMAP}
\end{figure}

%% file: Images.tex
\newpage

\section{Visual quantification hallucinations}\label{sec:imag}

The Figures below start from the same idealized TNG image, whose magnitude is artificially changed (second column). The two right-most images are the pixel-wise mean and variance ratio computed over 256 samples from the model applied on the second image. The Figures are corresponding to the values computed in Table~\ref{tab:hall}.

\begin{figure}[h!]
    \centering
    \begin{minipage}{0.9\textwidth}
        \centering
        \begin{subfigure}[b]{0.23\textwidth}
            \centering
            \includegraphics[width=0.895\linewidth]{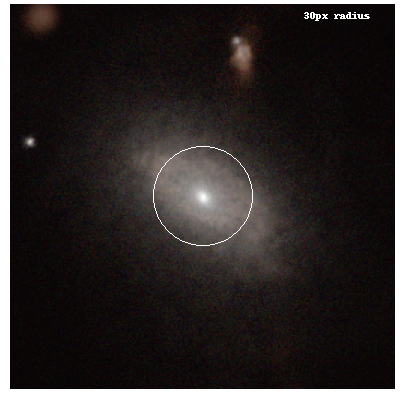}
            \vspace{0.05cm}
        \end{subfigure}
        \hfill
        \begin{subfigure}[b]{0.23\textwidth}
            \centering
            \includegraphics[width=0.895\linewidth]{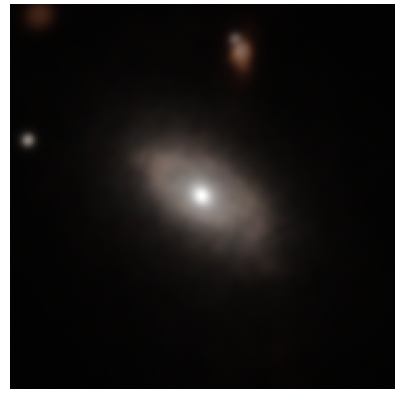}
            \vspace{0.05cm}
        \end{subfigure}
        \hfill
        \begin{subfigure}[b]{0.23\textwidth}
            \centering
            \includegraphics[width=0.9\linewidth]{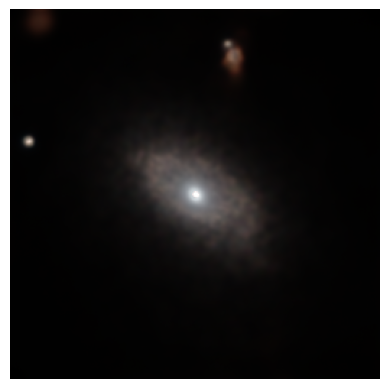}
        \end{subfigure}
        \hfill
        \begin{subfigure}[b]{0.25\textwidth}
            \centering
            \includegraphics[width=0.98\linewidth]{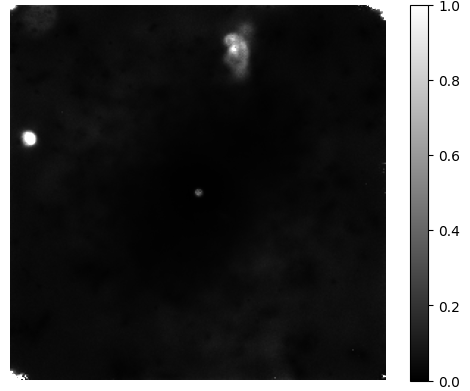}
        \end{subfigure}
    \end{minipage}
    \caption{\textit{Left to right}: Idealized - Simulated at \textbf{mag 18} - Pixel-wise mean - Variance ratio}
\end{figure}
\vspace{-0.5cm}
\begin{figure}[h!]
    \centering
    \begin{minipage}{0.9\textwidth}
        \centering
        \begin{subfigure}[b]{0.23\textwidth}
            \centering
            \includegraphics[width=0.875\linewidth]{test1_true.png}
            \vspace{0.05cm}
        \end{subfigure}
        \hfill
        \begin{subfigure}[b]{0.23\textwidth}
            \centering
            \includegraphics[width=0.895\linewidth]{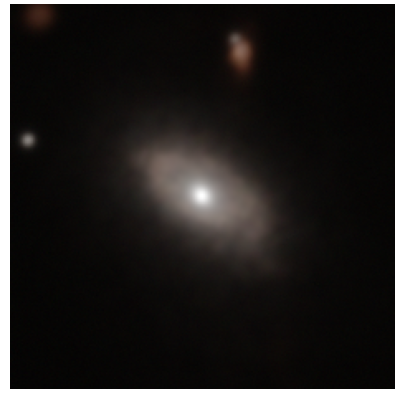}
            \vspace{0.05cm}
        \end{subfigure}
        \hfill
        \begin{subfigure}[b]{0.23\textwidth}
            \centering
            \includegraphics[width=0.9\linewidth]{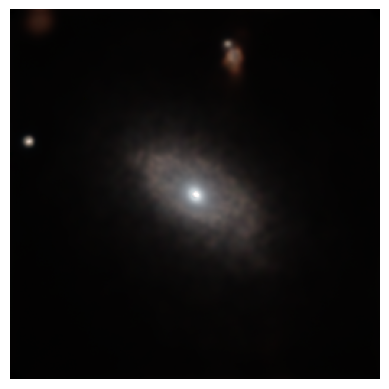}
        \end{subfigure}
        \hfill
        \begin{subfigure}[b]{0.25\textwidth}
            \centering
            \includegraphics[width=0.98\linewidth]{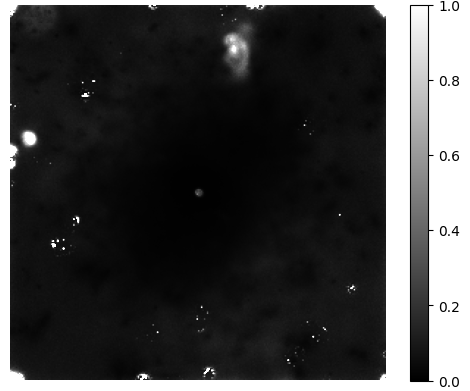}
        \end{subfigure}
    \end{minipage}
    \caption{\textit{Left to right}: Idealized - Simulated at \textbf{mag 19} - Pixel-wise mean - Variance ratio}
\end{figure}
\vspace{-0.5cm}
\begin{figure}[h!]
    \centering
    \begin{minipage}{0.9\textwidth}
        \centering
        \begin{subfigure}[b]{0.23\textwidth}
            \centering
            \includegraphics[width=0.875\linewidth]{test1_true.png}
            \vspace{0.05cm}
        \end{subfigure}
        \hfill
        \begin{subfigure}[b]{0.23\textwidth}
            \centering
            \includegraphics[width=0.895\linewidth]{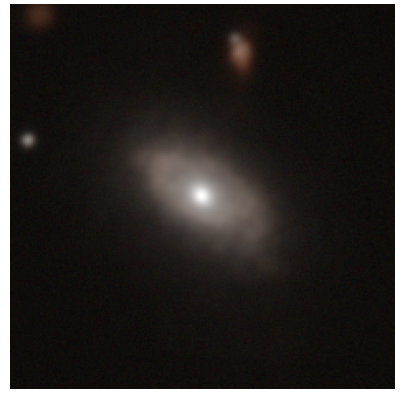}
            \vspace{0.05cm}
        \end{subfigure}
        \hfill
        \begin{subfigure}[b]{0.23\textwidth}
            \centering
            \includegraphics[width=0.9\linewidth]{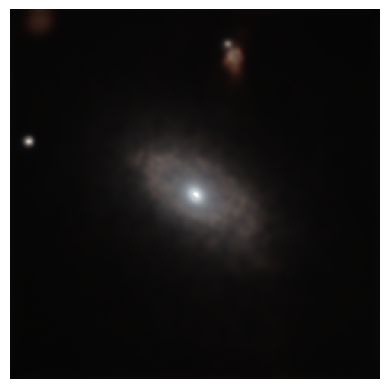}
        \end{subfigure}
        \hfill
        \begin{subfigure}[b]{0.25\textwidth}
            \centering
            \includegraphics[width=0.98\linewidth]{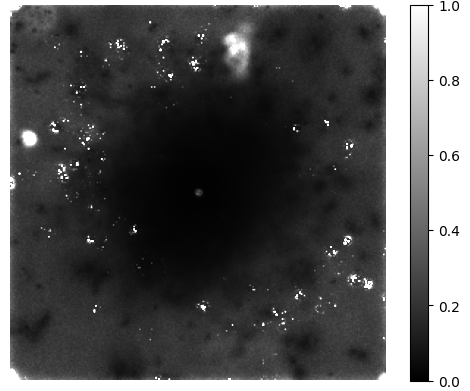}
        \end{subfigure}
    \end{minipage}
    \caption{\textit{Left to right}: Idealized - Simulated at \textbf{mag 20} - Pixel-wise mean - Variance ratio}
\end{figure}
\vspace{-0.5cm}
\begin{figure}[h!]
    \centering
    \begin{minipage}{0.9\textwidth}
        \centering
        \begin{subfigure}[b]{0.23\textwidth}
            \centering
            \includegraphics[width=0.875\linewidth]{test1_true.png}
            \vspace{0.05cm}
        \end{subfigure}
        \hfill
        \begin{subfigure}[b]{0.23\textwidth}
            \centering
            \includegraphics[width=0.895\linewidth]{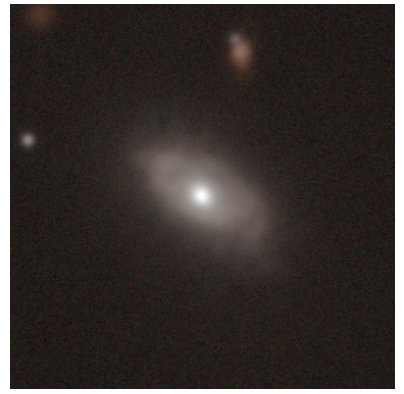}
            \vspace{0.05cm}
        \end{subfigure}
        \hfill
        \begin{subfigure}[b]{0.23\textwidth}
            \centering
            \includegraphics[width=0.9\linewidth]{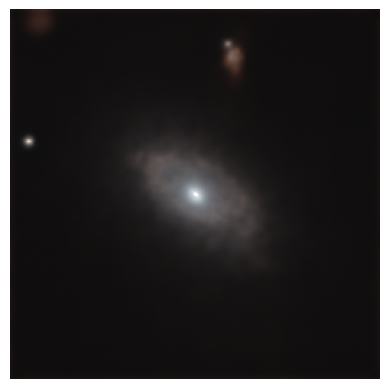}
        \end{subfigure}
        \hfill
        \begin{subfigure}[b]{0.25\textwidth}
            \centering
            \includegraphics[width=0.98\linewidth]{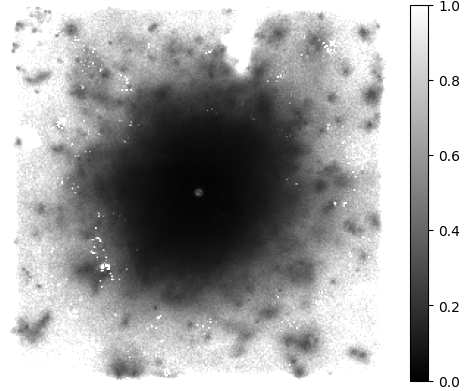}
        \end{subfigure}
    \end{minipage}
    \caption{\textit{Left to right}: Idealized - Simulated at \textbf{mag 21} - Pixel-wise mean - Variance ratio}
    
\end{figure}
\vspace{-0.5cm}
\begin{figure}[h!]
    \centering
    \begin{minipage}{0.9\textwidth}
        \centering
        \begin{subfigure}[b]{0.23\textwidth}
            \centering
            \includegraphics[width=0.875\linewidth]{test1_true.png}
            \vspace{0.05cm}
        \end{subfigure}
        \hfill
        \begin{subfigure}[b]{0.23\textwidth}
            \centering
            \includegraphics[width=0.895\linewidth]{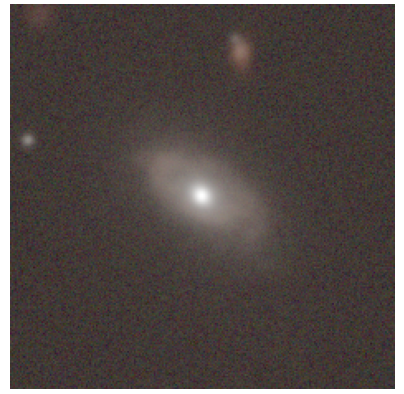}
            \vspace{0.05cm}
        \end{subfigure}
        \hfill
        \begin{subfigure}[b]{0.23\textwidth}
            \centering
            \includegraphics[width=0.9\linewidth]{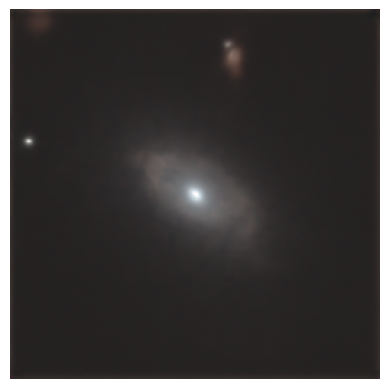}
        \end{subfigure}
        \hfill
        \begin{subfigure}[b]{0.25\textwidth}
            \centering
            \includegraphics[width=0.98\linewidth]{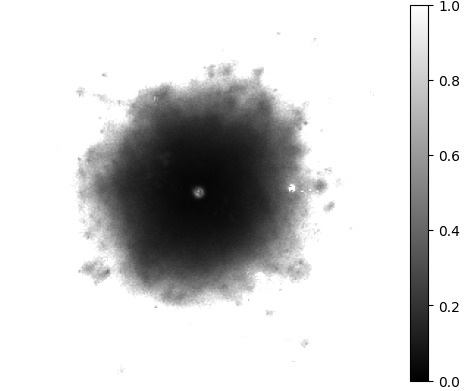}
        \end{subfigure}
    \end{minipage}
    \caption{\textit{Left to right}: Idealized - Simulated at \textbf{mag 22} - Pixel-wise mean - Variance ratio}
\end{figure}
\vspace{-0.5cm}
\begin{figure}[H]
    \centering
    \begin{minipage}{0.9\textwidth}
        \centering
        \begin{subfigure}[b]{0.23\textwidth}
            \centering
            \includegraphics[width=0.875\linewidth]{test1_true.png}
            \vspace{0.05cm}
        \end{subfigure}
        \hfill
        \begin{subfigure}[b]{0.23\textwidth}
            \centering
            \includegraphics[width=0.895\linewidth]{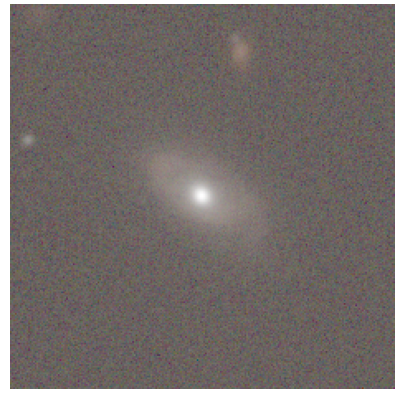}
            \vspace{0.05cm}
        \end{subfigure}
        \hfill
        \begin{subfigure}[b]{0.23\textwidth}
            \centering
            \includegraphics[width=0.9\linewidth]{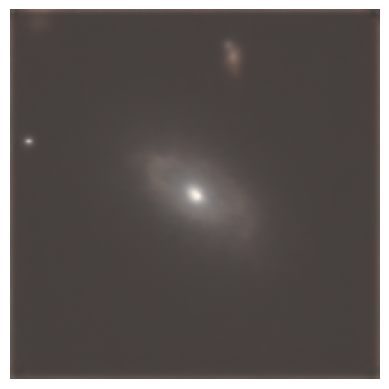}
        \end{subfigure}
        \hfill
        \begin{subfigure}[b]{0.25\textwidth}
            \centering
            \includegraphics[width=0.98\linewidth]{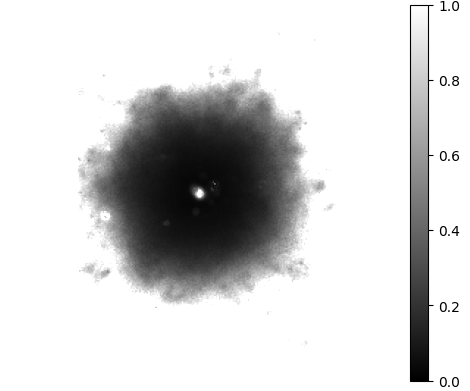}
        \end{subfigure}
    \end{minipage}
    \caption{\textit{Left to right}: Idealized - Simulated at \textbf{mag 23} - Pixel-wise mean - Variance ratio}
\end{figure}

%% file: main.bbl
\begin{thebibliography}{10}

\bibitem{Adam2022GalaxyPosterior}
A.~Adam, A.~Coogan, N.~Malkin, R.~Legin, L.~Perreault-Levasseur, Y.~D. Hezaveh, and Y.~Bengio.
\newblock Posterior samples of source galaxies in strong gravitational lenses with score-based priors.
\newblock {\em ArXiv}, abs/2211.03812, 2022.

\bibitem{adam2023echoesnoiseposteriorsamples}
A.~Adam, C.~Stone, C.~Bottrell, R.~Legin, Y.~Hezaveh, and L.~Perreault-Levasseur.
\newblock Echoes in the noise: Posterior samples of faint galaxy surface brightness profiles with score-based likelihoods and priors, 2023.

\bibitem{Aihara2021HSC}
H.~Aihara, Y.~AlSayyad, M.~Ando, R.~Armstrong, J.~F. Bosch, E.~Egami, H.~Furusawa, J.~Furusawa, S.~Harasawa, Y.~Harikane, B.~C. Hsieh, H.~Ikeda, K.~Ito, I.~Iwata, T.~Kodama, M.~Koike, M.~Kokubo, Y.~Komiyama, X.~Li, Y.~Liang, Y.-T. Lin, R.~H. Lupton, N.~B. Lust, L.~A. Macarthur, K.~Mawatari, S.~Mineo, H.~Miyatake, S.~Miyazaki, S.~More, T.~Morishima, H.~Murayama, K.~Nakajima, F.~Nakata, A.~J. Nishizawa, M.~Oguri, N.~Okabe, Y.~Okura, Y.~Ono, K.~Osato, M.~Ouchi, Y.-C. Pan, A.~A.~P. Malag'on, P.~A. Price, S.~L. Reed, E.~S. Rykoff, T.~Shibuya, M.~Simunovic, M.~A. Strauss, K.~Sugimori, Y.~Suto, N.~Suzuki, M.~Takada, Y.~Takagi, T.~Takata, S.~Takita, M.~Tanaka, S.~Tang, D.~S. Taranu, T.~Terai, Y.~Toba, E.~L. Turner, H.~Uchiyama, B.~Vijarnwannaluk, C.~Waters, Y.~Yamada, N.~Yamamoto, and T.~Yamashita.
\newblock Third data release of the hyper suprime-cam subaru strategic program.
\newblock 2021.

\bibitem{Anderson1982Reversetime}
B.~D.~O. Anderson.
\newblock Reverse-time diffusion equation models.
\newblock {\em Stochastic Processes and their Applications}, 12:313--326, 1982.

\bibitem{Bertero1998PSF3}
M.~Bertero, P.~Boccacci, and M.~Robberto.
\newblock Inversion method for the restoration of chopped and nodded images.
\newblock In {\em Astronomical Telescopes and Instrumentation}, 1998.

\bibitem{Bottrell2023IllustrisTNGHSC}
C.~Bottrell, H.~M. Yesuf, G.~Popping, K.~C. Omori, S.~Tang, X.~Ding, A.~Pillepich, D.~Nelson, L.~Eisert, H.~Gao, A.~D. Goulding, B.~S. Kalita, W.~Luo, J.~E. Greene, J.~Shi, and J.~D. Silverman.
\newblock Illustristng in the hsc-ssp: image data release and the major role of mini mergers as drivers of asymmetry and star formation.
\newblock 2023.

\bibitem{Chung2022}
H.~Chung, J.~Kim, M.~T. McCann, M.~L. Klasky, and J.~C. Ye.
\newblock Diffusion posterior sampling for general noisy inverse problems.
\newblock {\em ArXiv}, abs/2209.14687, 2022.

\bibitem{Dhariwal2021}
P.~Dhariwal and A.~Nichol.
\newblock Diffusion models beat gans on image synthesis.
\newblock {\em ArXiv}, abs/2105.05233, 2021.

\bibitem{Efron2011TweediesFA}
B.~Efron.
\newblock Tweedie’s formula and selection bias.
\newblock {\em Journal of the American Statistical Association}, 106:1602 -- 1614, 2011.

\bibitem{Gan_2021}
F.~K. {Gan}, K.~{Bekki}, and A.~{Hashemizadeh}.
\newblock {SeeingGAN: Galactic image deblurring with deep learning for better morphological classification of galaxies}.
\newblock {\em arXiv e-prints}, page arXiv:2103.09711, Mar. 2021.

\bibitem{Goodfellow2016}
I.~Goodfellow, Y.~Bengio, and A.~Courville.
\newblock {\em Deep Learning}.
\newblock MIT Press, 2016.
\newblock \url{http://www.deeplearningbook.org}.

\bibitem{Graikos2022DiffusionPlugP}
A.~Graikos, N.~Malkin, N.~Jojic, and D.~Samaras.
\newblock Diffusion models as plug-and-play priors.
\newblock {\em ArXiv}, abs/2206.09012, 2022.

\bibitem{Heusel2017FID}
M.~Heusel, H.~Ramsauer, T.~Unterthiner, B.~Nessler, and S.~Hochreiter.
\newblock Gans trained by a two time-scale update rule converge to a local nash equilibrium.
\newblock In {\em Neural Information Processing Systems}, 2017.

\bibitem{Ho2020Denoising}
J.~Ho, A.~Jain, and P.~Abbeel.
\newblock Denoising diffusion probabilistic models.
\newblock {\em ArXiv}, abs/2006.11239, 2020.

\bibitem{Kim2021Noise2ScoreTA}
K.~Kim and J.-C. Ye.
\newblock Noise2score: Tweedie's approach to self-supervised image denoising without clean images.
\newblock In {\em Neural Information Processing Systems}, 2021.

\bibitem{Lanusse2020DeepGM}
F.~Lanusse, R.~Mandelbaum, S.~Ravanbakhsh, C.-L. Li, P.~E. Freeman, and B.~P{\'o}czos.
\newblock Deep generative models for galaxy image simulations.
\newblock {\em Monthly Notices of the Royal Astronomical Society}, 504:5543--5555, 2020.

\bibitem{Lauritsen2021}
L.~Lauritsen, H.~Dickinson, J.~Bromley, S.~Serjeant, C.-F. Lim, Z.-K. Gao, and W.-H. Wang.
\newblock {Superresolving Herschel imaging: a proof of concept using Deep Neural Networks}.
\newblock {\em Monthly Notices of the Royal Astronomical Society}, 507(1):1546--1556, 07 2021.

\bibitem{Liu2007FaceHall}
C.~Liu, H.~Shum, and W.~T. Freeman.
\newblock Face hallucination: Theory and practice.
\newblock {\em International Journal of Computer Vision}, 75:115--134, 2007.

\bibitem{Lucy1994PSF2}
L.~B. Lucy.
\newblock Image restorations of high photometric quality.
\newblock In R.~J. Hanisch and R.~L. White, editors, {\em The Restoration of HST Images and Spectra - II}, page~79, January 1994.

\bibitem{McInnes2018UMAP}
L.~McInnes and J.~Healy.
\newblock Umap: Uniform manifold approximation and projection for dimension reduction.
\newblock {\em ArXiv}, abs/1802.03426, 2018.

\bibitem{Michalewicz2023PSF1}
K.~Michalewicz, M.~Millon, F.~Dux, and F.~Courbin.
\newblock Starred: a two-channel deconvolution method with starlet regularization.
\newblock {\em J. Open Source Softw.}, 8:5340, 2023.

\bibitem{Nelson2018TNG}
D.~Nelson, V.~Springel, A.~Pillepich, V.~Rodriguez-Gomez, P.~Torrey, S.~Genel, M.~Vogelsberger, R.~Pakmor, F.~Marinacci, R.~Weinberger, L.~Z. Kelley, M.~R. Lovell, B.~Diemer, and L.~E. Hernquist.
\newblock The illustristng simulations: public data release.
\newblock {\em Computational Astrophysics and Cosmology}, 6:1--29, 2018.

\bibitem{Nichol2021ImprovedDDPM}
A.~Nichol and P.~Dhariwal.
\newblock Improved denoising diffusion probabilistic models.
\newblock {\em ArXiv}, abs/2102.09672, 2021.

\bibitem{Nishizawa2020Photometric2}
A.~J. Nishizawa, B.~C. Hsieh, M.~Tanaka, and T.~Takata.
\newblock Photometric redshifts for the hyper suprime-cam subaru strategic program data release 2.
\newblock 2020.

\bibitem{Nishizawa2022Photometric3}
A.~J. Nishizawa, B.-C. Hsieh, M.~Tanaka, and the HSC~Collaboration.
\newblock Photometric redshifts for the hyper suprime-cam subaru strategic program data release 3.
\newblock {\em Publications of the Astronomical Society of Japan}, 74:247, 2022.
\newblock HSC PDR3 photometric data release.

\bibitem{Ntampaka2019TheRoleof}
M.~Ntampaka, C.~Avestruz, S.~Boada, J.~Caldeira, J.~Cisewski-Kehe, R.~D. Stefano, C.~Dvorkin, A.~E. Evrard, A.~Farahi, D.~P. Finkbeiner, S.~Genel, A.~A. Goodman, A.~D. Goulding, S.~Ho, A.~B. Kosowsky, P.~L. Plante, F.~Lanusse, M.~Lochner, R.~Mandelbaum, D.~Nagai, J.~A. Newman, B.~Nord, J.~E.~G. Peek, A.~Peel, B.~P{\'o}czos, M.~M. Rau, A.~Siemiginowska, D.~J. Sutherland, H.~Trac, and B.~D. Wandelt.
\newblock The role of machine learning in the next decade of cosmology.
\newblock {\em arXiv: Instrumentation and Methods for Astrophysics}, 2019.

\bibitem{Robbins1956AnEB}
H.~E. Robbins.
\newblock An empirical bayes approach to statistics.
\newblock 1956.

\bibitem{Ronneberger2015UNet}
O.~Ronneberger, P.~Fischer, and T.~Brox.
\newblock U-net: Convolutional networks for biomedical image segmentation.
\newblock {\em ArXiv}, abs/1505.04597, 2015.

\bibitem{Sampson2023SpottingHI}
M.~L. Sampson and P.~Melchior.
\newblock Spotting hallucinations in inverse problems with data-driven priors.
\newblock 2023.

\bibitem{Schawinski_2017}
K.~Schawinski, C.~Zhang, H.~Zhang, L.~Fowler, and G.~K. Santhanam.
\newblock Generative adversarial networks recover features in astrophysical images of galaxies beyond the deconvolution limit.
\newblock {\em Monthly Notices of the Royal Astronomical Society: Letters}, 467(1):L110–L114, Jan. 2017.

\bibitem{Scoville2006COSMOS}
N.~Scoville, H.~Aussel, M.~Brusa, P.~L. Capak, C.~M. Carollo, M.~Elvis, M.~Giavalisco, L.~Guzzo, G.~Hasinger, C.~D. Impey, J.-P. Kneib, O.~LeF{\`e}vre, S.~J. Lilly, B.~Mobasher, A.~Renzini, A.~Renzini, R.~M. Rich, D.~B. Sanders, E.~Schinnerer, E.~Schinnerer, D.~Schminovich, P.~Shopbell, Y.~Taniguchi, and N.~de~Grasse~Tyson.
\newblock The cosmic evolution survey (cosmos): Overview.
\newblock {\em The Astrophysical Journal Supplement Series}, 172:1 -- 8, 2006.

\bibitem{SmithDDPMgal}
M.~J. Smith, J.~E. Geach, R.~A. Jackson, N.~Arora, C.~Stone, and S.~Courteau.
\newblock {Realistic galaxy image simulation via score-based generative models}.
\newblock {\em Monthly Notices of the Royal Astronomical Society}, 511(2):1808--1818, 01 2022.

\bibitem{Song2020DenoisingDI}
J.~Song, C.~Meng, and S.~Ermon.
\newblock Denoising diffusion implicit models.
\newblock {\em ArXiv}, abs/2010.02502, 2020.

\bibitem{Song2019}
Y.~Song and S.~Ermon.
\newblock Generative modeling by estimating gradients of the data distribution.
\newblock In {\em Neural Information Processing Systems}, 2019.

\bibitem{Song2020improved}
Y.~Song and S.~Ermon.
\newblock Improved techniques for training score-based generative models.
\newblock {\em ArXiv}, abs/2006.09011, 2020.

\bibitem{Song2021SDE}
Y.~Song, J.~Sohl-Dickstein, D.~P. Kingma, A.~Kumar, S.~Ermon, and B.~Poole.
\newblock Score-based generative modeling through stochastic differential equations, 2021.

\bibitem{Starck2002DeconvolutionIA}
J.-L. Starck, E.~Pantin, and F.~Murtagh.
\newblock Deconvolution in astronomy: A review.
\newblock {\em Publications of the Astronomical Society of the Pacific}, 114:1051 -- 1069, 2002.

\bibitem{Stein1981EstimationOT}
C.~M. Stein.
\newblock Estimation of the mean of a multivariate normal distribution.
\newblock {\em Annals of Statistics}, 9:1135--1151, 1981.

\bibitem{Tanaka2017Photometric1}
M.~Tanaka, J.~Coupon, B.~C. Hsieh, S.~Mineo, A.~J. Nishizawa, J.~S. Speagle, H.~Furusawa, S.~Miyazaki, and H.~Murayama.
\newblock Photometric redshifts for hyper suprime-cam subaru strategic program data release 1.
\newblock {\em arXiv: Astrophysics of Galaxies}, 2017.

\bibitem{Venkatakrishnan2013PlugP}
S.~V. Venkatakrishnan, C.~A. Bouman, and B.~Wohlberg.
\newblock 5-29-2013 plug-and-play priors for model based reconstruction.
\newblock {\em IEEE}, 2013.

\bibitem{Vojtekova2020}
A.~Vojtekova, M.~Lieu, I.~Valtchanov, B.~Altieri, L.~Old, Q.~Chen, and F.~Hroch.
\newblock {Learning to denoise astronomical images with U-nets}.
\newblock {\em Monthly Notices of the Royal Astronomical Society}, 503(3):3204--3215, 11 2020.

\bibitem{Wang2013FaceHall2}
N.~Wang, D.~Tao, X.~Gao, X.~Li, and J.~Li.
\newblock A comprehensive survey to face hallucination.
\newblock {\em International Journal of Computer Vision}, 106:9--30, 2013.

\end{thebibliography}
